\documentstyle[11pt]{article}

\def\be{\begin{equation}}
\def\ee{\end{equation}}
\def\ba{\begin{eqnarray}}
\def\ea{\end{eqnarray}}
\def\de{\partial}
\begin{document}
\begin{titlepage}
\vskip0.5cm
\begin{flushright}
DFTT 8/95\\
hep-th/9501104\\
January 1995
\end{flushright}
\vskip0.8cm
\begin{center}
{\Large {\bf The 2D effective field theory of interfaces}}
\end{center}
\begin{center}
{\Large {\bf derived from 3D field theory}}
\end{center}
\vskip0.8cm
\begin{center}
{\large Paolo Provero \footnote{e--mail: provero@ge.infn.it}
 and Stefano Vinti \footnote{e--mail: vinti@to.infn.it}}\\
\vskip0.8cm
{\it Dipartimento di Fisica Teorica}\\
{\it Universit\`a di Torino}\\
{\it Via P. Giuria 1}\\
\it{10125 Torino, Italia}
\end{center}
\vskip1.5cm
\begin{abstract}
The one--loop determinant computed around the kink solution
in the 3D $\phi^4$ theory, in cylindrical geometry, allows one to obtain
the partition function of the interface separating coexisting phases.
The quantum fluctuations of the interface around its equilibrium
position are described by a $c=1$ two--dimensional conformal field
theory, namely a 2D free massless scalar field living on the interface.
In this way the capillary wave model conjecture for the
interface free energy in its gaussian approximation is proved.
\end{abstract}
\end{titlepage}

\section{Introduction}
The physics of interfaces separating coexisting phases in 3D systems
is dominated by long--wavelength, low--energy fluctuations; it is
therefore natural to describe the interface fluctuations in terms
of a 2D effective theory. A common assumption is to take the
interface free energy to be proportional to the area of the interface:
this is the well known capillary wave model (CWM) which is believed to
describe the interface physics~\cite{cwm,fgp}.
More recently, the CWM predictions  have been
made explicit one order beyond the gaussian approximation and verified
by means of numerical simulations on 3D spin systems to high accuracy
{}~\cite{cgv,cgvhp}.
\par
Despite these results, the CWM is an {\em ad hoc} 2D effective theory:
it is not known in general how to derive it from the
original 3D hamiltonians, except in the  zero--temperature limit
(see {\em e.g.}~\cite{Zia} and references therein).
\par
In this paper we provide an analytical derivation of the 2D effective
theory of interfaces in the framework of 3D Euclidean $\phi^4$ theory,
which is known to describe the scaling region of the Ising
model (for a general review see for instance~\cite{Parisi}).
Our result reproduces the predictions of the CWM in its gaussian
approximation: the partition function of an interface is proportional
to the partition function of a 2D, $c=1$ conformal field theory (CFT),
namely a free scalar field living on the interface.
\par
This result can be thought of as a new instance of dimensional reduction:
the relevant degrees of freedom of a physical system are described by
an effective theory of lower dimensionality.
\par
To be more precise, we consider the 3D $\phi^4$ theory in a cylindrical
geometry with two of the three space--time dimensions having finite
lengths $L_1$, $L_2$ and periodic boundary conditions.
Using $\zeta$--function regularization, we compute, in one--loop
approximation, the energy--gap $E(L_1,L_2)$ due to tunneling:
in the dilute--gas approximation, this quantity is proportional to
the partition function of an interface.
\par
The paper is organized as follows: in Sec.~2 we establish our
notations and review the expression of $E(L_1,L_2)$ in terms
of a functional determinant, regularized using the
$\zeta$--function method. In Sec.~3 we evaluate the determinant:
our main result is the expression (\ref{final}) for $E(L_1,L_2)$.
Sec.~4 is devoted to some concluding remarks.
\section{The interface partition function}
Consider the 3D field theory defined by the action
\be
S[\phi]=\int d^3x \left[\frac{1}{2}\de_\mu\phi\de_\mu\phi+
V(\phi)\right]
\ee
where
\be
V(\phi)=\frac{g}{4!}\left(\phi^2-v^2\right)^2
\ee
in Euclidean space--time with finite size in the "spatial" directions $x_i$
($i=1,2$) but infinite in the "time" direction $x_0$. We put periodic
boundary conditions on the finite sizes:
\be
\phi(x_0,x_1+L_1,x_2)=\phi(x_0,x_1,x_2+L_2)=\phi(x_0,x_1,x_2) \ \ .
\ee
The potential $V$ has two degenerate minima in $\phi=\pm v$ and a maximum
in $\phi=0$.
\par
A solution of the equations of motion connecting the two minima is
the kink
\be
\phi_{cl}(x)=v\tanh\left[\frac{m}{2}\left(x_0-a\right)\right]\label{kink}
\ee
where
\be
m=\left(\frac{gv^2}{3}\right)^{1/2}~~~,
\ee
and its action is
\be
S_c\equiv S[\phi_{cl}]=\frac{2m^3}{g}L_1 L_2\ \ .
\ee
The existence, in finite volume, of classical solutions connecting
the two degenerate minima of the potential, and hence of a non--vanishing
tunneling probability between the two minima, has the effect of
removing the double degeneracy of the vacuum which, in infinite
volume, is due to the spontaneous breaking of the $Z_2$ symmetry
$\phi\to -\phi$. The energy splitting is given, in one--loop
approximation, by (see {\em e.g.}~\cite{Coleman})
\be
E(L_1,L_2)=2 e^{-S_c}\left(\frac{S_c}{2\pi}\right)^{1/2}
\left|\frac{\det^\prime M}{\det M_0}\right|^{-1/2}\label{split}
\ee
where $M$ is the operator
\be
M=-\frac{\de}{\de x_\mu}\frac{\de}{\de x_\mu}
+V^{\prime\prime}(\phi_{cl}(x))~~~.
\ee
Here $\det^\prime$ indicates the determinant without the zero mode, which
is due to the freedom in choosing the kink location $a$, and
gives rise, when treated with the collective coordinates method,
to the prefactor $(S_c/2\pi)^{1/2}$. $M_0$ is the free--field
fluctuation operator
\be
M_0=-\de_\mu\de_\mu+m^2~~~.
\ee
\par
The computation of the energy splitting (\ref{split}) for the symmetric
case $L_1=L_2$ was done in Ref.~\cite{Mun1}.
We will see that the generalization of the calculation to asymmetric
geometries allows one to recognize the interface partition function
as the partition function of a $2D$ CFT.
\vskip0.5cm
We use $\zeta$--function regularization to compute the ratio of determinants
appearing in Eq.(\ref{split}). It is useful to express the operators
$M$ and $M_0$ as
\ba
&&M=Q(x_0)-\de_i\de_i\ \ \ \ \ (i=1,2)\\
&&M_0=Q_0(x_0)-\de_i\de_i
\ea
where
\ba
&&Q(x_0)=-\de_0^2+m^2-\frac{3}{2}m^2\frac{1}{\cosh^2\left[
\frac{m}{2}(x_0-a)\right]}\\
&&Q_0(x_0)=-\de_0^2+m^2~~~.
\ea
The regularized ratio of determinants appearing in Eq. (\ref{split}) is
then expressed as
\be
\frac{\det^\prime M}{\det M_0}=\exp\left\{-\left.\frac{d}{ds}\left[
\zeta_M(s)-\zeta_{M_0}(s)\right]\right|_{s=0}\right\}\label{det}
\ee
where the $\zeta$--function of an operator $A$ with eigenvalues $a_n$
is defined as
\be
\zeta_A(s)=\sum_n a_n^{-s}
\ee
The spectra of the operators $Q$, $Q_0$ and $-\de_i\de_i$ are known,
and the relevant $\zeta$--function is
\ba
\zeta_M(s)-\zeta_{M_0}(s)&=&{\sum_{n_1,n_2}}^\prime\left(\lambda_{n_1,n_2}
\right)^{-s}+\sum_{n_1,n_2}\left(\lambda_{n_1,n_2}+\frac{3}{4} m^2
\right)^{-s}\nonumber\\
&+&\sum_{n_1,n_2}
\int^{+\infty}_{-\infty} dp\  g(p)\left(\lambda_{n_1,n_2}+
p^2+m^2\right)^{-s}\label{zeta}
\ea
where the primed sum runs over $(n_1,n_2)\ne (0,0)$. Here $\lambda_{n_1
n_2}$ are the eigenvalues of the two--dimensional operator $-\de_i\de_i$
with periodic boundary conditions on the rectangle of sides $L_1$, $L_2$:
\be
\lambda_{n_1 n_2}=4\pi^2\left(\frac{n_1^2}{L_1^2}+\frac{n_2^2}{L_2^2}
\right)\ \ \ \ \ n_1,n_2\in Z~~~.
\ee
$g(p)$ is the difference between the spectral densities of $Q$ and
$Q_0$:
\be
g(p)=-\frac{m}{2\pi}\left(\frac{2}{p^2+m^2}+\frac
{1}{p^2+\frac{m^2}{4}}\right)~~~.
\ee
\par
\section{Evaluation of the determinant}
To complete our calculation we have to evaluate the $\zeta$--function
(\ref{zeta}). Following Refs.~\cite{Mun1,Mun2} we write
\be
\zeta_M(s)-\zeta_{M_0}(s)\equiv\zeta_1(s)+\zeta_2(s)
\ee
where
\ba
\zeta_1(s)&=&{\sum_{n_1,n_2}}^\prime\lambda_{n_1,n_2}^{-s}\\
\zeta_2(s)&=&\sum_{n_1,n_2}\left\{\left(\lambda_{n_1 n_2}
+\frac{3 m^2}{4}\right)^{-s}\right.\nonumber\\
&&\left.+\int^{+\infty}_{-\infty}
dp \ g(p)\left(\lambda_{n_1 n_2}+p^2+m^2\right)^{-s}\right\}
\ea
The term $\zeta_1$ can be recognized to be the $\zeta$--function of a
massless, 2D free scalar field on the rectangle of sides $L_1$, $L_2$
with periodic boundary conditions, {\it i.e.} on a torus~\cite{iz}.
{}From 2D CFT we know that its derivative in $s=0$ is~\cite{iz}
\be
\left.\frac{d\zeta_1}{ds}\right|_{s=0}=-2\log\left[L_1\left|\eta(\tau)
\right|^2\right]\label{zeta1}
\ee
where
\be
\tau\equiv i\frac{L_1}{L_2}
\ee
is modular parameter of the torus and $\eta(\tau)$ is the Dedekind
function.
When combined with the prefactor $(S_c/2\pi)^{1/2}$ coming from the
zero mode in Eq. (\ref{split}), this term produces precisely the
modular invariant partition function of the $c=1$ CFT defined by
a free massless scalar field.
\par
To evaluate $\zeta_2(s)$, we proceed like in Refs.~\cite{Mun1,Mun2}: we
write
\ba
\zeta_2(s)&=&\frac{1}{\Gamma(s)}\sum_{n_1 n_2}\int^{\infty}_{0} dt
\ t^{s-1}\left\{\exp
\left[-\left(\lambda_{n_1 n_2}+\frac{3 m^2}{4}\right)t\right]\right.
\nonumber\\
&&\left.+\int^{+\infty}_{-\infty} dp\  g(p)
\exp\left[-\left(\lambda_{n_1 n_2}+p^2+m^2\right) t\right]
\right\}
\ea
and, introducing the Jacobi theta function
\be
A(x)=\sum_{n}\exp\left(-\pi n^2 x\right)~~~,
\ee
we have
\ba
\zeta_2(s)&=&\frac{1}{\Gamma(s)}
\int^{\infty}_{0}dt\  t^{s-1} A\left(\frac{4\pi t}
{L_1^2}\right) A\left(\frac{4\pi t}{L_2^2}\right) F(m,t)\\
F(m,t)&=&\exp\left(-\frac{3}{4}m^2 t\right)+
\int^{+\infty}_{-\infty} dp\ g(p)
\exp\left[-\left(p^2+m^2\right)t\right]~~~.
\ea
Using Poisson's summation formula $A(x)$ is seen to satisfy
\be
A(x)=x^{-1/2}A(1/x)
\ee
which we use to express $\zeta_2$ as
\be
\zeta_2(s)=\zeta_2^{(a)}(s)+\zeta_2^{(b)}(s)
\ee
where
\ba
\zeta_2^{(a)}(s)&=&\frac{L_1 L_2}{4\pi}\frac{1}{\Gamma(s)}
\int^{\infty}_{0} dt\  t^{s-2}F(m,t)\\
\zeta_2^{(b)}(s)&=&\frac{L_1L_2}{4\pi}\frac{1}{\Gamma(s)}
\int^{\infty}_{0} dt\  t^{s-2}\left[A\left(\frac{L_1^2}{4\pi t}\right)
A\left(\frac{L_2^2}{4\pi t}\right)-1\right] F(m,t)
\ea
The term $\zeta_2^{(b)}$ is exponentially suppressed for large $L_1$,
$L_2$~\cite{Mun2} and will therefore be neglected in what follows.
$\zeta_2^{(a)}$ is then computed straightforwardly:
\ba
\zeta_2^{(a)}(s)&=&\frac{L_1L_2}{4\pi}\frac{1}{s-1}m^{2(1-s)}
\left\{\left(\frac{3}{4}\right)^{(1-s)}-\frac{3}{2\pi}\frac{\Gamma(1/2)
\Gamma(s-1/2)}{\Gamma(s)}\right.\nonumber\\
&&\left.-\frac{3}{8\pi}
\int^{+\infty}_{-\infty} dq\frac{(q^2+1)^{-s}}
{q^2+1/4}\right\}
\ea
so that
\be
\left.\frac{d\zeta_2^{(a)}}{ds}\right|_{s=0}=-\frac{3 m^2 L_1 L_2}
{4\pi}\left(1+\frac{1}{4}\log 3\right)\label{zeta2}~~~.
\ee
Therefore the $\zeta_2^{(a)}$ term provides simply a quantum correction
to the interface tension.
\par
Substituting (\ref{zeta1}) and (\ref{zeta2}) in (\ref{det}) and
(\ref{split}) we finally obtain
\be
E(L_1, L_2)=\frac{C}{\left[Im(\tau)\right]^{1/2}\left|\eta(\tau)\right|^2}
\exp(-\sigma L_1 L_2)\label{final}
\ee
where
\ba
&&C=\frac{2}{\sqrt{\pi}}\left(\frac{m^3}{g}\right)^{1/2}\\
&&\sigma=-\frac{2 m^3}{g}\left[1+\frac{3g}{16\pi m}\left(1+\frac{1}
{4}\log 3\right)\right]~~~.
\ea
\par
In Ref.~\cite{Mun1} the energy gap was computed in the symmetric
case $L_1=L_2$, in which the $\tau$--dependent contribution reduces
to a constant.
Notice that in~\cite{Mun1} the energy gap is expressed in terms
of the physical mass $m_{phys}$ (inverse of the correlation length)
and the renormalized coupling $u_R\equiv g_R/m_R$ where the renormalized
parameters $g_R$ and $m_R$ are defined according to a particular
renormalization scheme. However it is important to keep in mind that,
at one--loop, the renormalized parameters differ from the bare
ones by {\it finite} quantities: the one--loop Feynman diagrams
in 3D $\phi^4$ are finite after dimensional continuation. The
formulae needed to make contact between our result and the
one quoted in Ref.~\cite{Mun1} are
\ba
&&\frac{g}{m}\equiv u=u_R\left(1+\frac{31 u_R}{128\pi}+{\cal O}(u_R^2)\right)\\
&&m^2=m_{phys}^2\left[1+\frac{u_R}{16\pi}\left(-4+3\log 3\right)
+{\cal O}(u_R^2)\right]~~~.
\ea
\section{Conclusions}
The effective, long--wavelength 2D theory of interface fluctuations
in 3D $\phi^4$ theory has been derived from first principles by
analytical methods. The interface partition function turns out to
be proportional to the partition function of a free massless 2D
scalar field living on the interface.
In this way one is able to obtain a 2D conformal invariant field
theory by dimensional reduction of 3D field theory.
\par
This result is in agreement with the predictions of the capillary wave
model of interfaces, which {\it assumes} an interface free energy
proportional to the interface area. Indeed, the capillary wave model
in its gaussian approximation predicts exactly the functional form
(\ref{final}) for the interface partition function~\cite{Bunk,cgv}.
\par
The predictions of the capillary wave model were tested against
Monte Carlo simulations of spin systems in Refs.~\cite{cgv,cgvhp}.
In particular in~\cite{cgvhp} the model was successfully verified
{\it beyond} the gaussian approximation: it would be interesting
to investigate whether the CWM contributions beyond the gaussian
one can be derived in a field--theoretic framework.

\vskip0.5cm
{\bf Acknowledgements}
\vskip0.5cm
We are grateful to M. Caselle, F. Gliozzi and R. Levi for
useful discussions.

\end{document}